\documentclass[10pt,twocolumn,twoside]{IEEEtran} 

\usepackage{amsmath,algorithm,algorithmic, bm, graphicx}

\newcommand{\vp}{\mathbf{p}}
\newcommand{\vg}{\mathbf{g}}
\newcommand{\va}{\mathbf{a}}
\newcommand{\mG}{\mathbf{G}}
\newcommand{\mW}{\mathbf{W}}
\newcommand{\mS}{\mathbf{S}}
\newcommand{\valpha}{\bm{\alpha}}
\newcommand{\vbeta}{\bm{\beta}}

\DeclareMathOperator*{\argmin}{argmin}

\begin{document}

\title{Source localization in reverberant rooms using sparse modeling
and narrowband measurements}
\author{Gilles Chardon, Laurent Daudet,~\IEEEmembership{Senior Member,~IEEE}
\thanks{GC is with the Acoustics Research Institute, Austrian Academy of Sciences, Vienna, Austria. LD is with Institut Langevin, Paris Diderot University, Paris, France. This work has been submitted to the IEEE for possible publication. Copyright may be transferred without notice, after which this version may no longer be accessible}}

\maketitle

\begin{abstract}
We study two cases of acoustic source localization in a reverberant room, from a number of point-wise narrowband measurements. In the first case, the room is perfectly known. We show that using a sparse recovery algorithm with a dictionary of sources computed a priori requires measurements at multiple frequencies. Furthermore, we study the choice of frequencies for these measurements, and show that one should avoid the modal frequencies of the room. In the second case, when the shape and the boundary conditions of the room are unknown, we propose a model of the acoustical field based on the Vekua theory, still allowing the localization of sources, at the cost of an increased number of measurements. Numerical results  are given, using simple adaptations of standard sparse recovery methods.
\end{abstract}


 \begin{center} \bfseries EDICS Category: AUD-LMAP \end{center}

\IEEEpeerreviewmaketitle

\section{Introduction}

Acoustic or electromagnetic source localization is an inverse problem for which numerous
methods have been developed, based on various models and algorithms.
A common assumption is that a low number
of sources are present. This assumption can be modeled in various
ways, such as the low-dimensionality of a subspace build from
the measurements in subspace methods (e.g. MUSIC \cite{music}), or as the sparsity of
the measurements in a pre-defined dictionary. We will use here
this latter model, introduced for source localization by Malioutov et al.
\cite{malioutov}.

Another frequent assumption is that the wave propagation occurs
in free-field, i.e. that the acoustic field verifies the
Sommerfeld radiation condition. In this case, the Green function of the
medium is known, but this limits the range of application to
particular cases such as open environment or anechoic chambers, or at least to rooms with small reverberation.

The particular focus of this paper is the localization of a small number of sources in a reverberant environment, from frequency-domain measurements. There has been an increased activity in the last few years, where localization in this framework  was improved by explicitly taking into account the specifics of the environment, e.g. the shape of the room and the reflective properties of its walls, assuming these are known \cite{dokmanic, leroux}. The goal of this paper is to compare the performance of this approach with known environment, to a more generic model of reverberation based on a sparse model of the wavefield itself, that does not require the knowledge of the propagation environment. We first recall some existing methods for different
cases: known or unknown environment, and time or frequency domain.

In the case of known environments and in the time domain, by using the fact that the wave equation is invariant by reversing the time parameter, so-called time reversal techniques \cite{fink,Ing05} allow a robust localization of one source.  
After recording the sound radiated by a source on a array of
transducers, these recording are played backwards by the microphones.
The resulting soundfield (that can be either produced experimentally or simulated) focuses back to the location of the original source. However, its resolution is limited by the standard wave diffraction limit, and this method does not easily take into account prior knowledge on the source. 

Another method
in the known environment/time domain case is 
the use of cosparsity \cite{nam}. The soundfield created by a low number $J$ of sources
is solution to the wave equation with Neumann boundary conditions (in 
the ideal case of rigid walls)
$$
\left\{
\begin{array}{l}
\frac{\partial^2 u}{\partial t^2} -c ^2 \Delta u  = \sum_{j=1}^J s_j \delta_{x_j} \\
\frac{\partial u}{\partial n} = 0 \mbox { on } \partial\Omega_0
\end{array}
\right.
$$
where $c$ is the wave velocity, and $s_j$ the sound emitted by the $j$-th
source, located at $x_j$. After discretization (using e.g. finite differences), this can be interpreted as a cosparse model \cite{cosparse} for the pressure, allowing
the recovery of the positions of the sources and
the signals they emit.

In the frequency domain case, Dokmanic and Vetterli proposed a method allowing the localization
of sources in a known reverberant environment using measurements
in multiple frequency bands \cite{dokmanic}. They replaced the free-field impulse response
of the sources by the impulse response computed using the finite element
method, and used this as a dictionary. They used a multichannel Orthogonal Matching Pursuit (OMP)
to consider the different frequency bands. This method has interesting
properties, such as localization of sources without direct line of sight,
but as will be shown in this paper, needs more and more frequency bands
as the number of sources increases. A variant of this method in
the time domain was proposed by Le Roux et al. \cite{leroux}.

For the localization in unknown environment and using narrowband measurements,
we introduce a sparse model for soundfields based on the Vekua theory \cite{vekua, henricivekua},
and the associated sparse recovery algorithms. While this method needs more
measurements than the method introduced by Dokmanic and Vetterli,
it does not need any prior information on the shape
of the room or its boundary conditions, and can be used with only
one arbitrarily chosen frequency. For the sake of simplicity and
readability, most of the results will be given for propagation
in 2D domains. The adaptation to 3D domains is however straightforward.

The paper is structured as follows. We recall in section 2 the application
of sparsity to source localization in free-field using narrowband measurements.
We recall some results about Green functions in section 3.
In section 4 we recall the method developed by Dokmanic and Vetterli 
in a known reverberant, and study the necessity of multiple bands measurements.
In section 5, we introduce our sparse model for reverberated sources and the associated
algorithms  and we show corresponding numerical results.

\section{Sparsity and source localization}

Sparsity is a signal processing paradigm used in various domains,
such as compression, machine learning, inverse problems, etc.
The application to the particular problem of source localization
has been introduced by Malioutov et al. \cite{malioutov}. We give here the basic concepts
of sparsity applied to source localization.

The soundfield is measured on an array of $M$ microphones located
at points $x_m$,
and is assumed to be emitted by $N$ sources located at points $y_n$.
The pressure emitted by the $n$-th source measured at the $m$-th microphone
is given by $a_n G(y_n, x_m)$ where $a_n$ is the complex amplitude
of the $n$-th source, and $G$ the Green function of the medium. In a 2D free-field,
the Green function writes
\begin{equation}
G_0(y,x) = \frac{i}{4}H_0(k \| x - y\|).
\label{dec}
\end{equation}
Using $\vg(y) = (G_0(y, x_m))_m$, we can write the vector $\vp$
of measurements on the array as
$$\vp = \sum_n a_n \vg(y_n).$$
If we further assume that the points $y_n$ are located on a 
predefined grid of $L$ points $z_l$, we can build a dictionary $\mG$ of
possible sources, with $\vg(z_l)$ as the $l$-th column.
Using this dictionary, we can write
\begin{equation}
\vp = \mG \va
\label{pga}
\end{equation}
where $\va$ is a $L$-dimensional vector, with only $N$
non-zero coefficients whose indexes correspond to the positions of the sources.

The sparsity of $\va$ allows the use of sparse recovery methods,
although the number of pressure measurements is much lower than $L$ the dimension of the space ( equation \ref{pga} is underdetermined). More precisely, the compressed sensing paradigm indicates that, under some conditions on the measurements, the required number of measurements scales as N, the number of non-zero coefficients in a (i.e., the number of sources to recover), with $N << L$. Many algorithms exist for such sparse recovery problem, including Basis Pursuit \cite{bp}, or iterative algorithms (Matching Pursuit
(MP) \cite{mp}, Orthogonal Matching Pursuit (OMP) \cite{omp}, etc.). 

A critical point of this framework is the computation of the dictionary
and its numerical properties.
While the construction of the dictionary  is straightforward in the free-field case, its computation
in the case of a room is more involved, and the case of the unknown room requires a more complex model. Tools useful to the treatment of these two
cases are given in the next section.

\section{Properties of Green functions}
\label{sec_green}

In this section, we review some results on the Green functions.
These results will be used for the design of dictionaries for sparse source localization in reverberant environments.

\subsection{Green function in a closed room}

The Green function used in the previous section, in the
case of free-field propagation, is solution to the Helmholtz
equation with a point source as a right-hand side and with the
Sommerfeld radiation condition. This radiation condition
models the fact that the energy is radiating to infinity, and that
no energy is coming from the infinity to the sources.

In a closed room, this condition is replaced by boundary conditions,
usually expressed using the values of the pressure
and its derivatives at the boundaries. In the ideal case of
rigid walls, where there is no normal displacement of the air
relative to the walls, the boundary conditions used are the Dirichlet
boundary conditions. The normal displacement is proportional to the
normal derivative of the pressure, and the Green function
is solution to the boundary value problem
\begin{equation}
\left\{
\begin{array}{l}
\Delta G + k_0^2 G = \delta \\
\frac{\partial G}{\partial n} = 0 \mbox { on } \partial\Omega_0
\end{array}
\right.
\label{helm}
\end{equation}
In the general case, the solution (and thus, the dictionary of sources) cannot be obtained 
analytically, but we can still obtain decompositions that will be useful
for computing and modeling the Green functions.

The Green function (actually any solutions of (\ref{helm}) with
some constraints on the right hand side) can be expanded, in the $L_2$ sense, on the modal basis of the room. The vectors
of this basis are the eigenmodes of the Laplace operator
with Neumann boundary conditions, i.e. non-zero solutions
to
$$
\left\{
\begin{array}{l}
\Delta u + k^2 u = 0 \\
\frac{\partial u}{\partial n} = 0 \mbox { on } \partial\Omega_0
\end{array}
\right.
$$
The eigenmodes $u_n$ are associated to their spatial eigenfrequencies
$k_n$.

Using properties of these eigenmodes, it can be shown that the Green function has
the following modal expansion:
$$G(x, y) = \sum \frac{u_n(x) u_n(y)}{k^2 - k_n^2}.$$

The Green function can also be decomposed as the sum of a
particular solution of the Helmholtz equation with
right hand side, but no boundary conditions, and of a solution
to the homogeneous equation, such that the sum satisfies
the boundary conditions. A typical choice for the particular solutions
is the free-field Green function $G_0$. The homogeneous term $G_h$ is then
the solution to
$$
\left\{
\begin{array}{l}
\Delta G_h + k^2 G_h = 0 \\
\frac{\partial G_h}{\partial n} = -\frac{\partial G_0}{\partial n}  \mbox { on } \partial\Omega_0
\end{array}
\right.
$$
and the Green function writes
$$G = G_0 + g_h.$$

Numerous approximations schemes exist for the treatment
of solutions of the homogeneous Helmholtz equation
(Method of Fundamental Solutions \cite{mfs}, Boundary Element Methods \cite{bem}, etc).
We will here use the Vekua theory \cite{vekua, henricivekua}. A result of this theory
is that, in a 2D domain, a solution $u$ to the homogeneous Helmholtz equation be approximated by
sums of Fourier-Bessel functions:
$$u \approx \sum_{l = -L}^L \alpha_l J_l(kr) e^{in\theta}$$
or sum of plane waves:
$$u \approx \sum_{l = -L}^L \beta_l e^{i \vec k_l \cdot \vec x}$$
where the wavevectors  share their wavenumbers. Similar results are available
in 3D using spherical harmonics.

These decompositions are valid in any star-convex domain,
and do not depend on the boundary conditions. As the order needed
to achieve a good approximation is significantly lower than
e.g. the number of samples needed to apply the Shannon-Nyquist sampling theorem
to these quantities (linear with respect to $kL$ vs. quadratic),
these approximations provide a low-dimensional model for 
acoustical waves, in particular for the
homogeneous component $G_h$.

\section{Localization in a known room}

In the case of source localization in a reverberant room,
the Green function no longer has the simple form of the free-field case, but is dependent on the
shape of the room and on the boundary conditions. For
sake of simplicity, we will consider ideal rigid walls,
i.e. Neumann boundary conditions. To treat this case,
Dokmanic and Vetterli suggest to replace the free-field dictionary
by an ad-hoc dictionary computed a priori. The sound field
emitted by a source at location $y$ is solution to the
Helmholtz equation with a second right hand side and Neumann
boundary conditions:
$$
\left\{
\begin{array}{l}
\Delta p + k^2 p  = \delta_y \\
p  = 0 \mbox{ on } \partial \Omega
\end{array}
\right.
$$
By sampling these solutions at the positions of the measurements
and, as in the free-field case, sampling the domain where
the sources are assumed to be, we can build a dictionary
for the localization of the sources in this particular
room. Dokmanic and Vetterli combined this dictionary with
joint sparsity of the sources in different frequencies,
and a simple adaptation of the OMP algorithm.

\begin{figure}
\includegraphics[width=6cm]{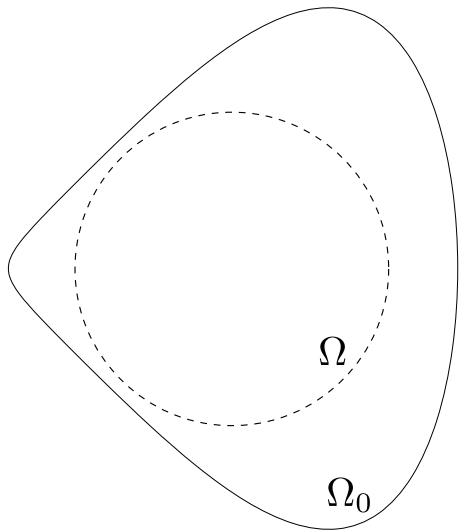}
\caption{Shape $\Omega_0$ of the room used for the numerical experiments.
$\Omega$ is the domain of interest for the unknown room case.}
\label{domain}
\end{figure}

We will show that, in this framework:
\begin{itemize}
\item it is actually necessary to use measurements
at multiple frequencies
to locate more that one source,
\item the choice of the frequencies is critical.
\end{itemize}

We first analyze how the reverberant case relates to the free-field case.
As pointed out above, the Green function
can be decomposed as a free-field term $G_0$ added to
a homogeneous term $G_h$.
Near the modal frequencies, that is when a nonzero solution
with zero normal derivative on the border exists
and the relative power of $G_h$ increases, this corrective
term can be much larger than the free-field contribution,
which changes the properties of the dictionary.

A more precise characterization of this perturbation
can be obtained using the modal basis. As
pointed out in section \ref{sec_green}, the Green function
can be decomposed as
$$G(x,y) = \sum \frac{u_n(x) u_n(y)}{k^2 - k_n^2}$$

Near some eigenfrequency $k_m$, the mode $u_n$ (if the eigenfrequency
is non-degenerate) will
dominate the Green function, which be almost constant
with respect to the position of the source. A consequence 
on the dictionary is that, if at least a measurement is not on
a nodal line, its coherence (i.e. the maximal scalar product
between columns) tends to 1 as the frequency approaches
an eigenfrequency. This is detrimental to the localization,
as the coherence of the dictionary has to be low
to ensure reconstruction by a sparse recovery algorithm such as OMP \cite{Tropp}.
In our case, it means that it is in practice impossible to locate
more that one source in this case.

Measuring the impulse response at frequencies near
an eigenfrequency will provide
no information on the location of the sources, and between two 
eigenfrequencies, the Green function will be dominated by the
two modes associated to these frequencies. This limits
the amount of information available at a given frequency,
making it necessary to use multiple frequencies, which are not close
to an eigenfrequency.

\begin{figure}
\includegraphics[width=8cm]{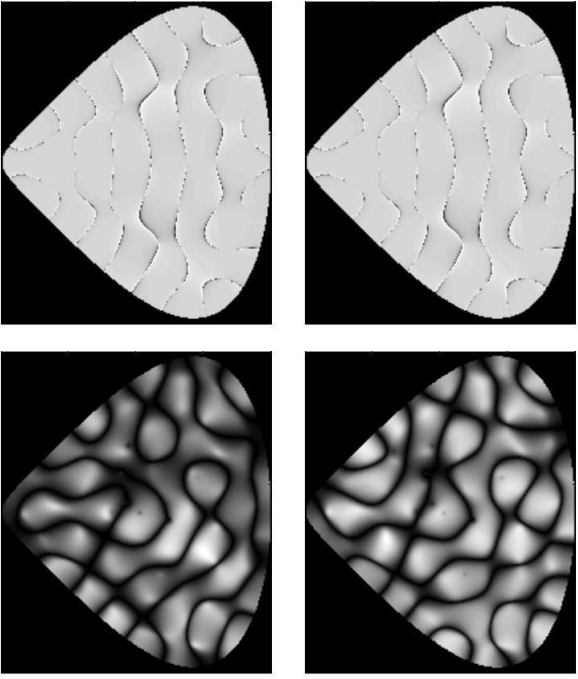}
\caption{Known room - Correlations between the dictionary and the measurements for
2 different sources, at an eigenfrequency (top), and between eigenfrequencies (bottom).}
\label{correlm}
\end{figure}

\begin{figure}
\includegraphics[width=8cm]{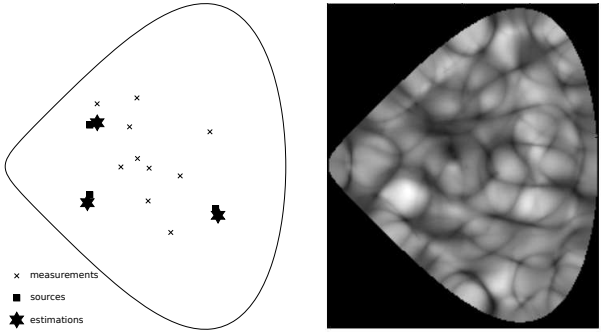}
\caption{Known room - Left: localization for three sources with 10 measurements and 6 frequencies.
Right: correlations at the first step.}
\label{omp3}
\end{figure}

To illustrate this, we simulate the propagation in the room pictured on
figure \ref{domain}, described by the parametric equations
\begin{equation}
\left\{
\begin{array}{l}
x= \cos t \\
y= \sin t + \frac{1}{3}\sin 2t
\end{array}
\right.
\ \ \ \ t \in [0, 2\pi)
\end{equation}
The dimensions of this room are approximately 2$\times$2.3.
We plot on figure \ref{correlm} the correlations between
the dictionary and the signals emitted by two different sources, at two different frequencies:
an eigenfrequency of the room, and a frequency at mid-point between eigenfrequencies.
For the eigenfrequency, the correlations for the two different sources 
are very similar, while for the other case, the correlations are different,
but do not exhibit a clear maximum allowing the identification of the sources.
Using multiple frequencies allows a clear localization. This is visible on
figure \ref{omp3} that shows the correlations between the dictionary and
the measurements for the first step of OMP, using 10 measurements and 6 frequencies, in order to locate 3 sources. A clear maximum is visible at the location of the most powerful source,
and all 3 sources are then identified in an iterative way.

In order to evaluate the localization problem in a known room, comprehensive simulations are run with varying numbers of sources,
microphones, frequencies, and three choice of frequencies:
\begin{itemize}
\item random draw of frequencies within a given interval.
\item modal frequencies of the room
\item means of two successive modal frequencies
\end{itemize}

We use the FreeFem++ software to simulate the data, and
a least-square method based on (\ref{dec}) and the Vekua approximations
to compute the dictionary \cite{eisenstat}. The modal frequencies
are computed using the Method of Particular Solutions \cite{betcke}.
At each trial, the positions of the measurements and of the sources are drawn randomly inside 
the 2-dimensional domain $\Omega_0$  pictured on figure \ref{domain}.

\begin{figure}
\includegraphics[width=8cm]{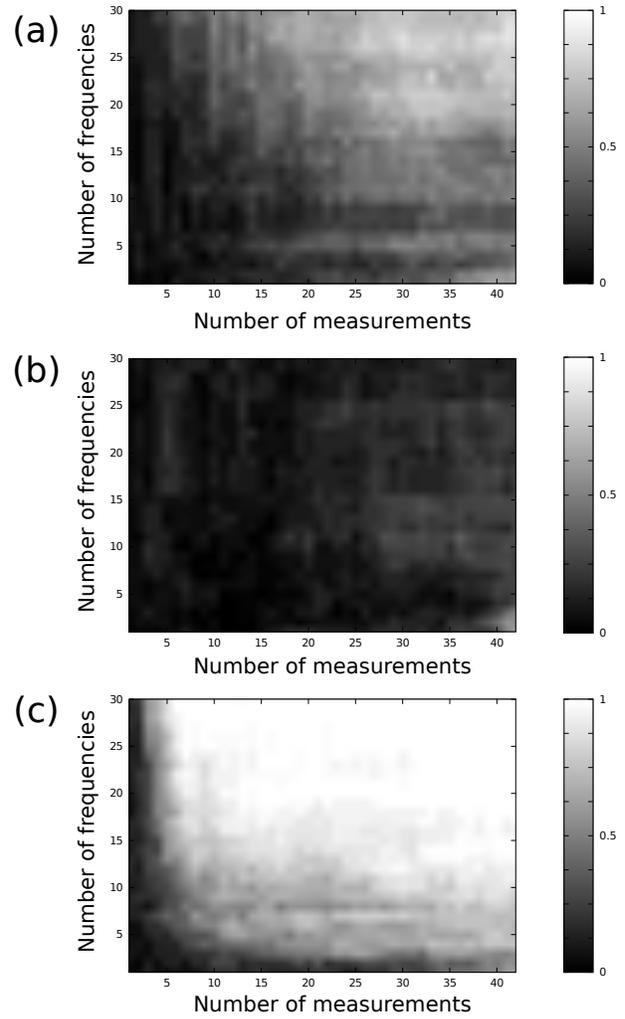}
\caption{Known room - Probability of successful localization for the three different choices of
frequencies,
with 2 sources, for 20 trials as a function of the number of
measurements and the number of frequencies. a) random draw of the frequencies, b) modal frequencies of the room, c) means of two successive modal frequencies.
Black = no source localized, white = all sources localized, $\epsilon = 0.2$.}
\label{cmpf}
\end{figure}

Localization results for 2 sources are shown on figure \ref{cmpf}.
The experiment is repeated 20 times, for a number of measurements
from 1 to 42, and a number of frequencies from 1 to 31. A source
is considered localized if an estimated source is at a distance less than
$\epsilon = 0.2$. The respective performances of the three choices of frequencies
are clearly different, and coherent with our analysis.
These results highlight clear differences in the three possible strategies for choosing the frequencies: while the random choice has mediocre performance
and the use of the modal frequencies does not yield exploitable results,
using frequencies between these modal frequencies makes
the localization possible, although a large number of 
frequencies is needed to achieve robust localization. Even in this case, using too few frequencies
prevents the localization of the sources. At least 10 frequencies are needed to recover
the two sources with a success rate of at least 80\%. 
We see that both a minimal number of measurements, as well as a minimal number
of well chosen frequencies are needed to localize sources in
a known reverberant room.

\section{Unknown room}

We known turn to the case of an unknown room,
for which we do not know the shape and/or the boundary conditions.
This makes the direct problem, and thus the computation of the dictionary, impossible. We propose
an alternative sparse modelization of the soundfield
radiated by sources, which allows localization
from narrowband measurements in unknown rooms using simple adaptations of
sparse recovery algorithms.

\subsection{Sparse modeling of the soundfield}

We assume that the sources and the measurements are located in
a domain of interest $\Omega$, contained in the domain of the
room $\Omega_0$ (see figure \ref{domain}).
The pressure $p$ radiated by the sources is solution to the Helmholtz
equation with a right hand side, and boundary conditions which are unknown
to us. We can however decompose $p$ as the sum of a particular solution
to the Helmholtz equation $p_s$, arbitrarily chosen, and a solution
to the homogeneous equation $p_0$.
 
Using the Vekua theory, and although the boundary conditions
are unknown to us, the homogeneous component $p_0$ can be approximated by sums of Fourier-Bessel
functions or plane waves:
$$p_0 \approx \sum_{l = -L}^L \alpha_l J_l(kr) e^{in\theta}
\mbox{ or } p_0 \approx \sum_{l = -L}^L \beta_l e^{i \vec k_l \cdot \vec x}$$
where the wavevectors $\vec k_l$ are uniformly sampled
on the circle or the sphere of radius $k$.
Note that sources in the room, but outside the domain of interest $\Omega$
will be included in this component. This can be viewed both as a feature
of the method, which is able to neglect unwanted sources outside of
its domain of interest, but also as a disadvantage, as it makes
localization of sources possible only inside the convex hull
of the sampling points.

Although a natural choice
for $p_s$ is to use the free-field Green function, we use its real part only, i.e. the Bessel function of second kind $Y_0$: The pressure produced by $N$ sources at positions $y_n$ with
complex amplitudes $a_n$ is given by:
$$p_s = \sum_{n=1}^N  \frac{i a_n}{4} Y_0(k \| x - y_n \|).$$
Indeed, the imaginary part, a Bessel function of the first kind,
can be included in the homogeneous component. This also emphasizes
that we cannot really locate sources, but more precisely right
hand sides of the Helmholtz equation.
The component $p_s$
has a sparse decomposition in a dictionary similar to the
one introduced in section 2.

After sampling at the location of the measurements, the model writes
\begin{equation}
\vp \approx \mS\valpha + \mW\vbeta
\label{sparsedec}
\end{equation}
where $\mS$ is a dictionary of sources, with $S_{kl} = Y_0(k \| x_k - y_l\|)$ and $\mW$ a dictionary of Fourier-Bessel functions
$W_{kl} = J_l(k \|x_k\|)$ or plane waves $W_{kl} = \exp(i \vec k_n \cdot \vec x_k)$.
$\valpha$ is the vector containing the coefficients of the sources
(assumed to be sparse),
and $\vbeta$ the coefficients of the Fourier-Bessel functions
(not sparse, but low-dimensional).

\subsection{Algorithms}

We propose two numerical  methods to localize sources using this sparse model.

The first is based on Orthogonal Matching Pursuit. In standard OMP,
the sources would be localized by correlating the measurements
with the source dictionary $\mS$. Here however, these correlations
would be corrupted by the homogeneous term $p_0$. To avoid
this, the measurements are projected on the
orthogonal of the space spanned by the columns of $\mW$.
At each iterations, the residual is obtained
by projecting the measurement vector $\vp$ on the space spanned by $\mW$ and the estimated sources.
This scheme is equivalent to using a dictionary consisting
of the union of $\mW$ and $\mS$, and forcing the first steps of OMP
to choose the atoms of $\mW$, or to using standard OMP after projecting the
dictionary of sources on the orthogonal complement of $\mW$.

This algorithm takes as inputs a dictionary of sources, a dictionary
used to approximate the homogeneous field (e.g. plane waves or Fourier-Bessel functions), the measurements, and a stopping criterion (e.g. number of sources to be localized).

\begin{algorithm}
\caption{Greedy source localization algorithm}
\label{algoOMP}
\begin{algorithmic}
\REQUIRE measurements $\mathbf p$, number of sources $n$, plane waves
or Fourier-Bessel dictionary $\mathbf W$,
source atoms $\mathbf s_{\mathbf y}$
\ENSURE estimated positions of the sources $\mathbf y_j$

$\mathbf p_s \leftarrow  \mathbf p - \mathbf W \mathbf W^\dagger \mathbf p$

\FOR{$j = 1$ to $n$}
\STATE $\mathbf y_j \leftarrow \max_{\mathbf{y}} |\left< \mathbf s_{\mathbf y}, \mathbf p_s \right>|$
\STATE $\mathbf W \leftarrow \left( \mathbf W | \mathbf s_{\mathbf y_j} \right)$
\STATE $\mathbf p_s \leftarrow  \mathbf p-\mathbf W \mathbf W^\dagger \mathbf p$
\ENDFOR
\end{algorithmic}
\end{algorithm}

The second method is a variant of Basis Pursuit. In equation
\ref{sparsedec}, the vector $\bm{\alpha}$, containing the activations of the sources, is assumed to be sparse,
while no assumption is made on the vector $\bm{\beta}$ of coefficients of the decomposition of
the homogeneous field. To recover the location of the sources, we minimize
the $\ell_1$-norm of  $\bm{\alpha}$, promoting its sparsity,
added to the $\ell_2$-norm of $\bm{\beta}$:
\begin{equation}
(\hat{\bm\alpha}, \hat{\bm\beta}) = \argmin_{(\bm\alpha, \bm\beta)}
\|\bm\alpha\|_1 + \|\bm\beta\|_2 \mbox{ s.t. } \|\vp - 
\mS  \bm\alpha - \mW  \bm\beta\| \leq \epsilon
\label{l1min}
\end{equation}
where $\epsilon$ is the expected noise level. This minimization problem
is a particular case of the Group Basis Pursuit problem.

While we here use measurements at only one frequency, which is sufficient in this
case, both these methods can be extended to the case of multiple frequencies,
through the use of an adaptation of OMP to treat multiple frequencies,
or a mixed norm for the minimization based method.

\subsection{Numerical results}

We test the OMP-based method for various frequencies and numbers of sources.
For each case, we estimate the probability of localizing the sources
as a function of the number of measurements and the number of Fourier-Bessel
functions used to approximate the homogeneous field. We assume that a source is successfully localized if one estimated source lies within a tolerance region of radius $\epsilon = 0.2$.

We use the same domain as above, but restrict the domain of interest to
a disk $\Omega$ of diameter $R=1.4$, in which the sources are randomly drawn.

\paragraph{Distribution of the samples}

We first test three different sampling strategies, for which the sampling
points are drawn using three different probability densities:
\begin{itemize}
\item uniform density in the domain,
\item uniform density on its border,
\item 50\% in the domain, 50\% on its border.
\end{itemize}

\begin{figure}
\includegraphics[width=8cm]{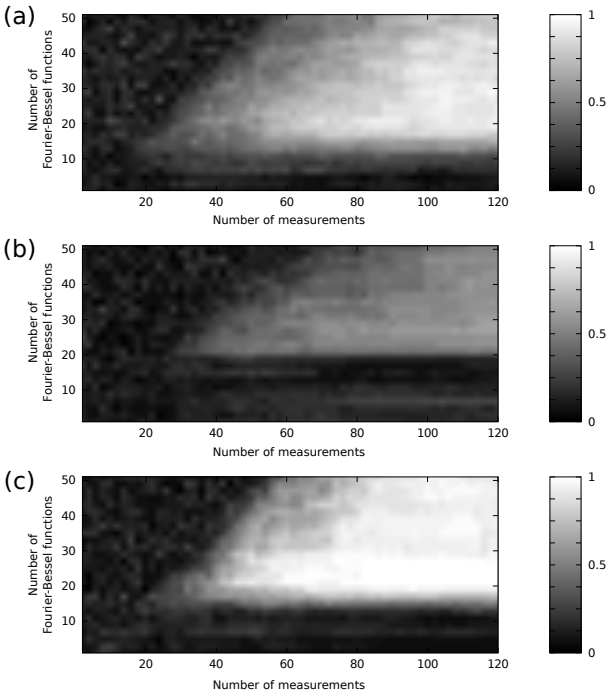}
\caption{Unknown room - Probability of successful localization for the three different sampling densities,
with 2 sources and $k = 10$, for 20 trials as a function of the number of
measurements and the number of Fourier-Bessel functions. a) uniform density in the domain,
b) uniform density on its border, c) mixture between these densities.
Black = no source localized, white = all sources localized, $\epsilon = 0.2$.}
\label{cmpspl}
\end{figure}

Results for these 3 densities are given on figure \ref{cmpspl}, for the case of
two sources with $k = 10$.

Sampling only on the border fails, as it is actually impossible
to distinguish the field created by a source from a homogeneous field
using only measurements on the border. Indeed, if one or more sources radiate
a pressure $p_s$ on the border, the solution to
$$
\left\{
\begin{array}{l}
\Delta p_0 + k^2 p_0  = 0 \\
p_0 = p_s \mbox{ on } \partial \Omega
\end{array}
\right.
$$
is an homogeneous field with the same value on the border of $\Omega$.

Mixed sampling has slightly better performances than
interior sampling, in particular for high numbers of Fourier-Bessel
functions. This is likely due to the fact the higher order Fourier-Bessel
functions are better identified with mixed sampling.

\paragraph{Number of measurements and Fourier-Bessel functions}

\begin{figure}
\includegraphics[width=8cm]{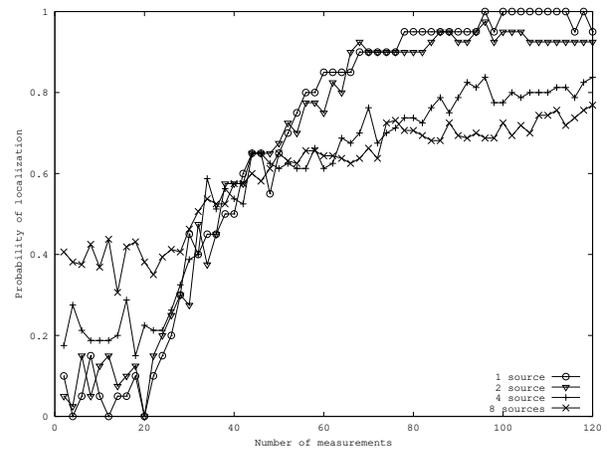}
\caption{Unknown room - Probability of successful localization with 
varying number of sources and number of measurements.
$k = 10$, 21 Fourier-Bessel functions.}
\label{cmpnsource}
\end{figure}

\begin{figure}
\includegraphics[width=8cm]{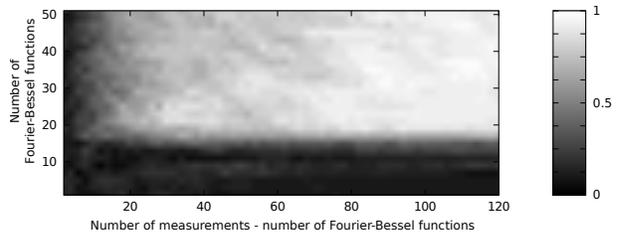}
\caption{Unknown room - Probability of successful localization with 2 sources 
at $k=10$, as a function of the number of measurements minus the number of
Fourier-Bessel functions.}
\label{cmpc}
\end{figure}

\begin{figure}
\includegraphics[width=8cm]{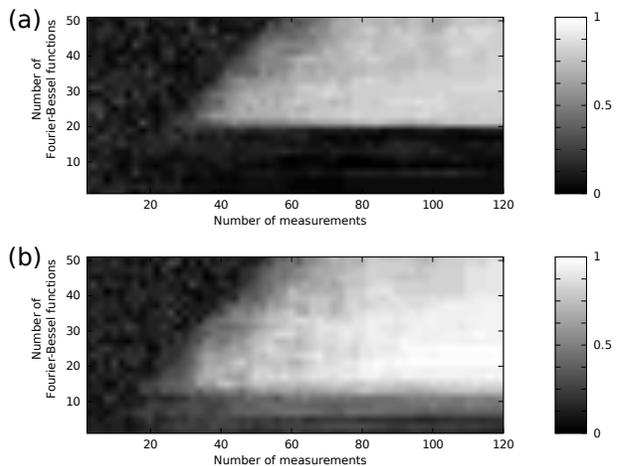}
\caption{Unknown room - Probability of successful localization with 2 sources 
for a modal frequency ($k = 9.98$) and a frequency between two modes
($k = 10.08$)}
\label{cmpmode}
\end{figure}

\begin{figure}
\includegraphics[width=8cm]{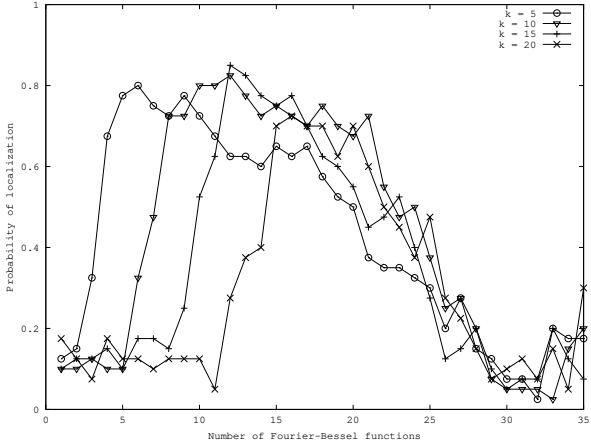}
\caption{Unknown room - Probability of successful localization with 2 sources 
for $k=5,10,15,20$, 60 measurements and various numbers of Fourier-Bessel functions.}
\label{cmpnfb}
\end{figure}

We here explain the particular shape of the domain of parameters for which the method works.

On figure \ref{cmpnsource}, we compare the performance for varying number
of sources and measurements with $k=10$ and $N_{fb} = 21$. When $N_m < N_{fb}$,
the localization fails (the growing percentages of localization as the number of sources increases
when $N_m < N_{fb}$ is an artefact of the way we measure the performance of the method, as more and more sources get localized due to pure chance).

On figure \ref{cmpc}, we show the performance of the method
with the same parameters as on figure \ref{cmpspl}-c, but
as a function of the number of measurements minus the number of Fourier-Bessel
functions instead of the number of measurements. This figure shows that as long
as there are enough Fourier-Bessel functions to capture the reverberant part of
the acoustic field, the performance of the method is more or less independent of
the number of Fourier-Bessel functions.

\paragraph{Wavenumber}

In the known room case, the particular structure of the dictionary
makes the localization possible only when using multiple frequencies
that were not modal frequencies of the room. The proposed method
for the unknown room case is less sensitive to the frequency of the measurements.
Results
of the proposed methods are given on figure \ref{cmpmode} for two different frequencies,
an eigenfrequency and a frequency between two modes. 
These frequencies are chosen close enough so that the main difference in behavior is due
to their modal or not character, and not to their respective magnitude.
The main difference between the two
cases is that a larger number of Fourier-Bessel functions has to be used
to capture the homogeneous part in the case of a modal frequency. This is expected as the homogeneous part has more energy in this case, and
high-order components that were not large enough to perturb the localization
have to be eliminated. This obviously makes the minimal number of measurements
higher, but no unreasonably so.
While some differences
can be seen figure \ref{cmpmode}, the overall performance is similar, if slightly better
for the non-modal case. While the choice of the frequency is not as critical as in the case of the known room, it is still preferable to use frequencies between
modes to locate sources in this case.

We test, for different wavenumbers ($k=5,10,15,20$) and fixed number of
measurements (60) and sources (2), the performance of the method
as a function of the number of Fourier-Bessel functions.
As seen on figure \ref{cmpnfb} the minimal number of Fourier-Bessel functions required
to localize sources  depends on the wavenumber. This
minimal number is approximately $kD$ where $D$ is the diameter of $\Omega_0$.
Using too many Fourier-Bessel functions is detrimental to the localization,
as projecting on the orthogonal of the space
spanned by these functions diminishes the effective number of measurements
available for the localization.

\paragraph{Using multiple frequencies}

\begin{figure}
\includegraphics[width=8cm]{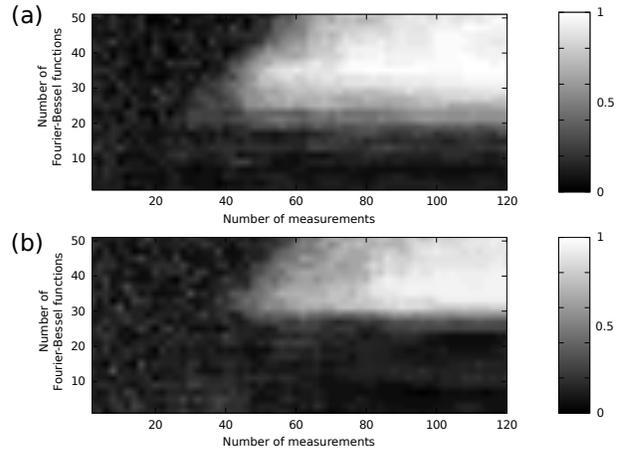}
\caption{Unknown room -  a) Probability of successful localization with 2 sources,
using multiple frequencies ($k=10$, $k=15$, and $k=20$).
b) Results for $k=20$ only.}
\label{hyper}
\end{figure}

On figure \ref{hyper}, we show the results of localization using
multiple frequencies ($k=10$, $k=15$, and $k=20$), and
compare it to the results of the method with only $k=20$.
When enough Fourier-Bessel functions are used to capture
the reverberant field at all the frequencies, using
multiple frequencies gives better results that using only one. In particular,
less measurements are needed to ensure a good probability of localization.
When too few Fourier-Bessel functions are used
for the highest frequency, but enough for the lowest ones (i.e.
approximately between 20 and 30 Fourier-Bessel functions) the results are
slightly better as the estimation succeeds for the low frequencies,
but are not reliable as the measurements at the highest frequency
are perturbed by the reverberant field.

\begin{figure}
\includegraphics[width=6cm]{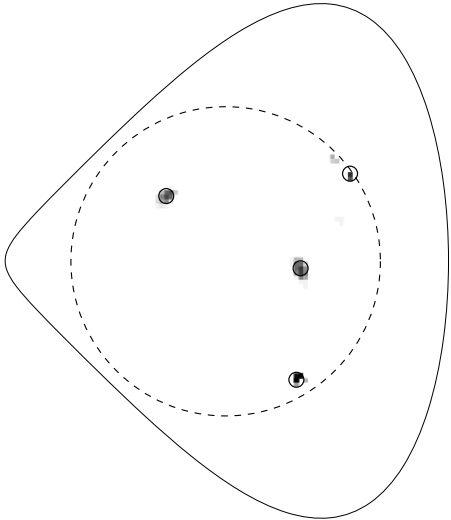}
\caption{Unknown room - Result of the $l_1$-minimization scheme. Power of estimated sources in grey level.
True positions are marked
with circles.}
\label{l1}
\end{figure}

\paragraph{Basis Pursuit based method}
To close this section, we give an example of localization of 4 sources
using the $\ell_1$-minimization scheme introduced above (implemented
using SPGL1 \cite{spgl1paper, spgl1}).
In this example,
we localize 4 sources, with 50 measurements, at $k = 10$ and
21 Fourier-Bessel functions. The output of the $l_1$ minimization
is drawn on figure \ref{l1}, along with the considered domain and
the domain of propagation.

\section{Conclusion}

Our experiments confirm that narrowband localization of sources in known or unknown reverberant rooms
is possible using adequate models. However, the two cases have quite different requirements
in terms of measurements. While the known room case can deal with a small number of measurements,
the numerical properties of the dictionaries require the use of multiple and carefully chosen frequencies. More precisely, we show that the measurements should be done between the modal frequencies of the room.

Localizing sources in a unknown reverberant environment is possible using only one frequency,
but needs a larger number of measurements, used to separate the direct response from the reverberation.
Another difference is that this scheme can only localize sources inside the convex hull of the antenna.

\section*{Acknowledgments}

GC is supported by the Austrian Science Fund (FWF) START-project FLAME (“Frames and Linear Operators for Acoustical Modeling and Parameter Estimation”; Y 551-N13).
LD is on a joint affiliation with Institut Universitaire de France, and is supported by LABEX WIFI under references ANR-10-LABX-24 and ANR-10-IDEX-0001-02 PSL*.
GC performed part of this work at Institut Langevin.

\bibliographystyle{ieeetr}
\bibliography{source}

\end{document}